\documentstyle[aps,twocolumn,floats,epsf]{revtex}
\begin{document}
\draft
\title{Fabrication and characterization of metallic nanowires}
\author{C. Untiedt, G. Rubio, S. Vieira and N. Agra\"{\i}t}
\address{Instituto Universitario de Ciencia de Materiales\\``Nicol\'{a}s Cabrera'',\\
Laboratorio de Bajas Temperaturas, \\ Departamento de F\'{\i}sica de la
Materia Condensada, C-III, \\ Universidad Aut\'{o}noma de Madrid, 28049
Madrid, Spain}
\date{March 25, 1997}
\maketitle

\begin{abstract}
The shape of metallic constrictions of nanoscopic dimensions (necks)
formed using a scanning tunneling microscope (STM) is shown to depend
on the fabrication procedure.
Submitting the neck to repeated plastic deformation cycles makes possible
to obtain long necks or nanowires. Point-contact spectroscopy results
show that these long necks are quite crystalline, indicating that the
repeated cycles of plastic deformation act as a ``mechanical annealing''
of the neck.
\end{abstract}
\pacs{61.16.Ch, 73.40.Cg, 62.20.Fe}

\par
\vspace{0.5cm}


\section{Introduction}

The development of new experimental techniques such as the mechanically
controllable break junctions (MCBJ) and the scanning tunneling microscope
(STM) has made possible the formation and study of atomic-size junctions or
contacts between macroscopic metals. Both techniques allow for a very
precise control of the relative position of two electrodes which is the key
to the study of these nanojunctions. These related techniques differ in the
way the nanojunction is formed: in the case of the STM the tip is driven
into the substrate to form a large contact and then it is pulled back, while
in the MCBJ a fine wire is pulled until it fractures and the two fragments
are brought back into contact.

Superconducting and normal contacts have been extensively studied at liquid
helium temperature (4.2 K) \cite{Muller,Agrait,Rodrigo,Rodrigo2,Smith} and
normal contacts at room temperature \cite{Pascual,Olesen}. More recently the
mechanical properties and their correlation with the conductance have been
studied using a combination of force and tunneling microscopy techniques
\cite{PRL1,Durig,PRL2}.

Theoretically, the atomistic mechanisms during plastic deformation of the
contact \cite{Landman,Lynden-Bell,Landman2} and their effect on the
conductance \cite{Todorov} have been investigated using molecular dynamics
(MD) simulations and a tight-binding approximation, and different geometries
have been considered by conductance calculations using a free electron
approximation \cite{Bogachek,Torres1,Torres2,Bratkovsky}.

In spite of all this effort no direct experimental geometrical and
structural information of these necks has been presented. In this
article we present a simple model that makes possible to obtain an estimate
of the shape of the necks and their evolution during plastic
deformation. We will also show that the type of neck produced depends on
its deformation history and that it is possible to fabricate {\em long}
necks or {\em nanowires.} These nanowires are quite crystalline as
evidenced by their electronic transport properties at low temperatures.

\begin{figure}[h]
\vspace{10mm}
\begin{center}
\leavevmode
    \epsfxsize=70mm
 \epsfbox{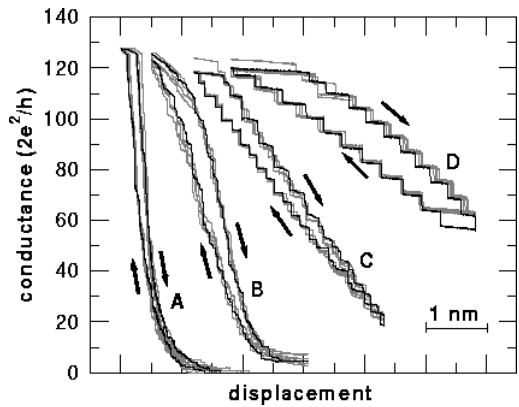}
 \end{center}
\vspace{0in}

\caption[]{Four different sets of experimental conductance {\em vs}
relative tip-substrate displacement cycles for Au at 4.2 K.
Each set consists of 5 consecutive elongation-contraction cycles
(one of the cycles is colored black while the other four are gray
for clarity). The arrows indicate the direction of motion.}
\label{fig.curves}
\end{figure}

\section{Determination of the geometry of necks}

In STM and MCBJ experiments on nanojunctions
the current that traverses the contact at a small
constant bias voltage (several milivolts) is measured as a function of
the relative displacement of the electrodes. Before a metallic contact is
established the electronic current is due to the tunneling effect and the
dependence of the current with distance is exponential \cite
{Gimzewski,Rodrigo}. Metallic contact is signaled by
an abrupt increase of the current
(the conductance for this first contact is of the order
of $2e^2/h$, where $e$ is the electron charge,
and $h$ is Planck's constant).
The variation of the current with the relative displacement of
the electrodes depends on
how the junction deforms plastically and can be quite diverse as attested by
Fig.\ \ref{fig.curves}, where we show four different sets of STM
experimental curves.

In this section we will first discuss the conductance of an ideal
metallic constriction and its relation to the minimal
cross-sectional area, and then we will present a simple slab model
for the evolution of the constriction with relative displacement
which makes possible to obtain an estimate of the shape of the
constriction and mechanical properties of the neck.

\subsection{Conductance of a metallic constriction}

Electronic transport through a metallic constriction whose dimensions are
smaller than the mean free path of the electrons \cite{mfpAu} is ballistic.
The conductance of such a constriction was first derived semiclassically by
Sharvin \cite{Sharvin}. More recently Torres {\em et al.}\cite{Torres1}
using an exact quantum calculation derived a corrected version of Sharvin
formula,
\begin{equation}
\label{eq.Sharvin}G_\infty \approx \frac{2e^2}h\left( \pi \frac A{\lambda
_F^2}-b\frac P{\lambda _F}\right)
\end{equation}
where $\lambda _F$ is the Fermi wavelength;
$A$ and $P$ are the area and perimeter of the minimal cross section;
$b$ is a parameter that depends on the shape of the neck,
being $1/2$ for a cylindrical tube ($\theta =0$), and $1/4$ for a
circular hole ($\theta =\pi /2$).
The classical Sharvin prediction \cite{Sharvin,Jansen}
corresponds to $b=0$. As the contact area increases the correction due to
the perimeter becomes relatively less important (about 5\% for a contact
radius of 2 nm for Au). A more realistic approximation to a real
constriction that takes into account the work function  of the material
instead of a hard wall potential \cite{Mole} reduces this correction even
further.

Equation \ref{eq.Sharvin} can be used to estimate the minimal cross section
of the constriction. Note that if the neck is ballistic and not too small
the error introduced by not knowing the aperture angle is negligible.
However the existence of defects or disorder would have the effect of
decreasing the conductance for a given geometry and using eq.\ \ref
{eq.Sharvin} would underestimate the neck diameter. These effects have been
studied by Todorov and coworkers using a molecular dynamics simulation
to obtain the evolution of the atomic positions during plastic deformation
of the neck and a  tight-binding model for the conductance corresponding to a
given atomic configuration \cite{Todorov}.
These authors find a reduction in the conductance of up to 20\% with respect
to the free electron result. Rough boundaries \cite{Bratkovsky} also affect
the oscillating structure due to the opening and closing of conductance
channels but
the smooth part of the conductance is still well described by the modified
Sharvin formula. More extended defects would cause an even larger reduction
\cite{Serena}.

Experimental information on the degree of disorder in the constriction can
be obtained using point-contact spectroscopy \cite{Jansen}. The derivative
of the differential conductance of a point contact (PC) contains information
about the inelastic electron backscattering. For ballistic point contacts
the dominant inelastic scattering mechanism is phonon scattering, and the
Eliashberg function for the electron-phonon interaction in the point contact
situation, or point contact spectroscopic (PCS) curve, is given by \cite
{Ralph}
\begin{equation}
\label{eq.spectral}\alpha^{2} F_{p}=-\frac{3}{32\sqrt{2}}\frac{h^{3/2}
k_{F}^{2}}{4 \pi^{2} m} R^{3/2} \frac{\mbox{d}^2 I}{\mbox{d}V^2}.
\end{equation}
where $R=\mbox{d} V/\mbox{d}I$, is the differential resistance; $k_{F}$ is
the Fermi wave number; and $m$ is the electron mass. The amplitude of the
phonon induced-peaks is reduced if there is elastic scattering (for example,
due to impurities or defects) in the contact region. Consequently, a large
PCS amplitude indicates that the constriction is indeed ballistic.

\subsection{Slab model}

\label{subsec.evolution}

\begin{figure}
\vspace{0mm}
\begin{center}
\leavevmode
    \epsfxsize=70mm
 \epsfbox{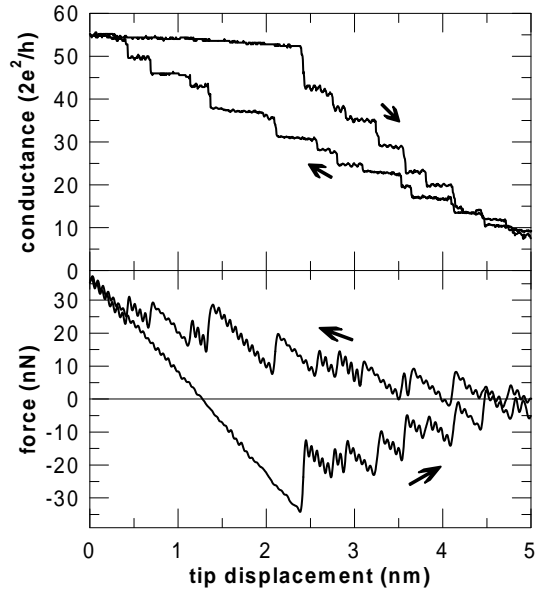}
 \end{center}
\vspace{0in}

\caption[]{Simultaneous measurement of conductance and force during
an elongation-contraction  cycle, for Au at 300 K.
The arrows indicate the direction of motion.}
\label{fig.forcecurve}
\end{figure}

A simple slab model, similar to that used by Torres {\em et al.} \cite
{MoleSlab} can be used to relate the changes in the minimal cross section as
the neck length is varied to the shape of the neck. This model is suggested
by the results of STM/AFM experiments \cite{PRL2}. In these experiments, in
which the conductance and force as a function of relative displacement are
measured simultaneously, it is observed (see Fig.\ \ref{fig.forcecurve})
that when the conductance is rather constant the force varies linearly,
while the abrupt jumps in conductance are correlated to abrupt force
relaxations. Between the relaxations deformation is elastic (no energy is
dissipated). This correlation has been clearly demonstrated in
nanometer-sized contacts (several hundreds of conductance
quantum units) \cite{PRL1} and
in atomic-sized contacts (just a few quantum units) \cite{PRL2}. Note that
changes in conductance due to effects like conductance quantization would
never show as abrupt jumps, but rather as smooth structure in the steps \cite
{MoleSlab}. This behavior has a clear interpretation: the neck deforms
elastically until the stress reaches a critical value
(about an order of magnitude larger than the bulk value \cite{PRL1,PRL2}),
and then  relaxes abruptly to a new configuration.

\begin{figure}[h]
\vspace{0mm}
\begin{center}
\leavevmode
    \epsfxsize=70mm
 \epsfbox{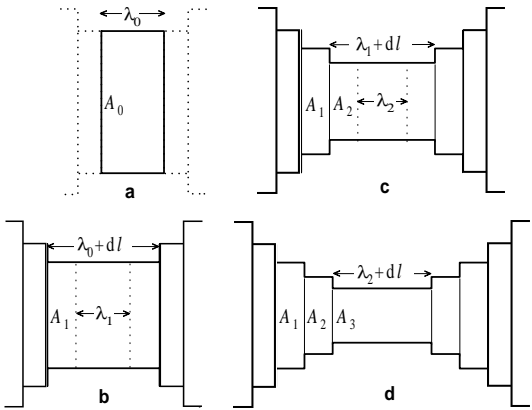}
 \end{center}
\vspace{0in}

\caption[]{Plastic deformation of a neck consisting
of cylindrical slabs. From one configuration to the next
only the central slab elongates, while the rest of the neck
does not change.}
\label{fig.model}
\end{figure}

Let us model the neck as a constriction with cylindrical geometry,
consisting of slabs of different radii and thicknesses, symmetrical with
respect to a plane normal to its narrowest section as depicted in
Fig.\ \ref{fig.model}a. The elastic properties of the neck
{\em e.g.} the Young's modulus $E$ and Poisson's ratio $\mu $
are identical to the bulk  values (this is reasonable from a
theoretical  viewpoint and has also been shown experimentally
in refs.\ \onlinecite{PRL1,PRL2}).
The basic assumption for this model is
that only the narrowest part of the neck deforms plastically.
This is likely to be the case because the weakest spot will
become the narrowest cross section even if this was not initially the case
(due for example to the existence of a very defective spot).
This assumption could break down for temperatures larger than
about 50\% of the  melting temperature for which diffusion will be
important.

When  the neck is being pulled the central slab will change to a longer
(and narrower due to volume conservation) configuration in order
to relax the stress.
For instance if the central slab has initially a cross
section $A_0=\pi R_0^2$, and a portion $\lambda _0$ of its length is
involved in the change of configuration, the new cross section would be $%
A_1=A_0\lambda _0/(\lambda _0+\Delta l)$, where $\Delta l$ is the change in
length (see Fig.\ \ref{fig.model}b), and the number of slabs increases by
two. After stretching $n$ intervals (assuming the changes in length are always
equal to $\Delta l$), the increase in length would be $n\Delta l$, the
minimal cross section
\begin{equation}
\label{eq.crosssection}A_n=\frac{A_{n-1}\lambda _{n-1}}{\lambda
_{n-1}+\Delta l},
\end{equation}
and the spring constant $k_n$ of the whole neck
\begin{equation}
\frac 1{k_n}=\frac 1E\left( \sum_{i=1}^{n-1}\frac{\lambda _{i-1}+\Delta
l-\lambda _i}{A_i}+\frac{\lambda _{n-1}+\Delta l}{A_n}\right) +\frac 1{k_0},
\end{equation}
where $k_0$ is the spring constant of the neck before starting the
elongation. The central portion of the neck consists of slabs of cross
section $A_i=A_{i-1}\lambda _{i-1}/(\lambda _{i-1}+\Delta l)$, and thickness
$(\lambda _{i-1}+\Delta l-\lambda _i)/2$. Note that we only know the shape
of the pulled portion of the neck but the initial neck is unknown.
This parameter $\lambda$ which we may call {\em plastic deformation length,} is
related to portion of the neck that changes plastically and
in general it will depend on the cross
section, length and history of the neck.
In the limit $\Delta l \rightarrow 0$, the plastic deformation length
can be written as  $\lambda=-(\mbox{d}\ln A/\mbox{d}l)^{-1}$.
Note that $\lambda$ is a phenomenological parameter
that will be determine from the experimental data
using eq.\ \ref{eq.crosssection}.
We will  see that the plastic deformation
length varies during plastic deformation of a given neck, and depending also
on the fabrication procedure.

\begin{figure}
\vspace{0mm}
\begin{center}
\leavevmode
    \epsfxsize=70mm
 \epsfbox{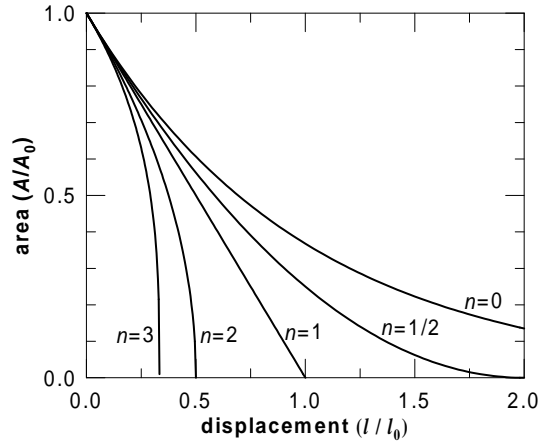}
 \end{center}
\vspace{0in}
\caption[]{Evolution of the conductance for different power law dependences
of the plastic deformation length $\lambda$ as given by eq.\ \ref{eq.power1}
and \ref{eq.power2}
}
\label{fig.evolution}
\end{figure}


As an illustration of how the plastic deformation length reflects on the
conductance curves, let us consider
that the plastic deformation length at
each step depends only the cross section
$\lambda=\alpha_0 A^{n}$, the differential equation for $A$
can be solved analytically. For an initial elongation zero we have
\begin{equation}
A=\left[A_0^n - \frac{l n}{\alpha_0}\right]^{1/n}, \hspace{1cm}%
\mbox{for $n\not=0$},
\label{eq.power1}
\end{equation}
\begin{equation}
A=A_0 \exp(-l/\lambda_0),\hspace{1cm} \mbox{for $n=0$,}
\label{eq.power2}
\end{equation}
where $l$ is the elongation.

\begin{figure}[h]
\vspace{0mm}
\begin{center}
\leavevmode
    \epsfxsize=70mm
 \epsfbox{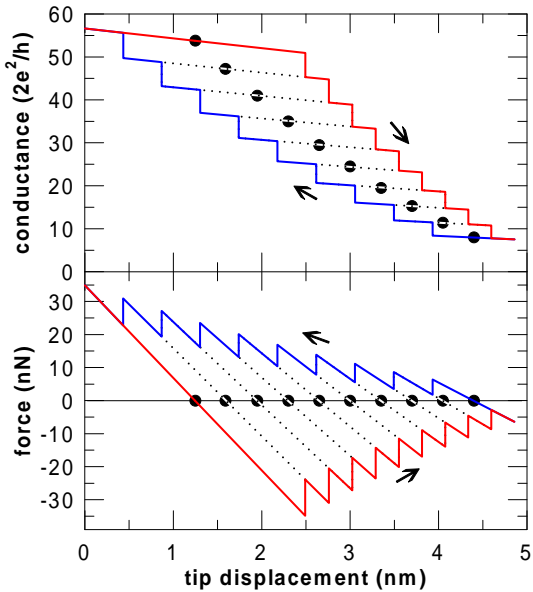}
 \end{center}
\vspace{0in}
\caption[]{Simulation using the slab model of the conductance and force during
an elongation-contraction  cycle. The relaxed position for
each configuration is marked with a black dot.
We are depicting a fairly ideal case in which the
configurations during elongation are
the {\em same} as the configurations during contraction.}
\label{fig.forcemodel}
\end{figure}

The evolutions of the minimal cross-section, and consequently of
the conductance, with the changes in length are shown in
Fig.\ref{fig.evolution}. In Ref. \onlinecite{Durig}, Stalder {\em
et al.\ } \cite{Durig} considered only the case \mbox{$n=0$}. Fig.\
\ref{fig.forcemodel} shows a simulation of the neck evolution
including the effect of elastic deformation. The black dots
represent the equilibrium (relaxed) configurations, and we have
assumed that the relaxations occur at a fixed value of the apparent
pressure (force divided by minimal cross-section area) as observed
experimentally \cite{PRL1,PRL2}.

\section{Experimental results}

\subsection{Fabrication of the necks}

The necks studied in this article have been obtained using an STM and an
STM/AFM combination, and the experiments where performed in different
conditions: low temperatures (4.2 K) in He exchange gas, room temperature in
He gas, and room temperature and ambient conditions. The fabrication
procedure described in this section works with ductile metals like Pb, Au,
Al and Sn. In this article we will only present the results for Au.

The basic procedure consists of crashing a clean metallic tip into a clean
metallic substrate in a controlled manner. The STM is convenient because
besides the high degree of control on tip-substrate separation (vertical
positioning), it makes possible the selection of different spots on the
substrate (horizontal positioning). We normally use the same metal for both
tip and sample because this guarantees that the composition of the neck is
homogeneous. As the tip is pressed against the substrate both electrodes
deform plastically and then bind by cohesive forces forming a metal contact.
Retraction of the tip results in the formation of a connective neck that
elongates plastically and eventually breaks. In all the ductile metals
studied rupture takes place through a gradual decrease of the cross section
of the neck. Measuring the current through the neck at a fixed bias voltage
of 10-100 mV applied between tip and substrate makes possible to know the
conductance as a function of the displacement of the tip relative to the
substrate and follow the evolution of the neck. As will be shown, this
evolution depends on the history of the contact.

In order to form a metal contact, tip and substrate surface must be clean.
Nevertheless, the preparation procedure is not very critical: it is
necessary to get rid of oxides and organic contaminants. For Au samples
sonication in acetone or cleaning in a 3:1 concentrated H$_{2}$SO$_{4}$:30\%
H$_{2}$O$_{2}$ solution work fine. For Pb or Al scratching the surfaces with
a clean blade is adequate. Tips are always clipped and care must be taken
not to produce a very fine tip that could bend elastically since we want all
the deformation (both plastic and elastic) to take place at the contact. It
must be emphasized that the actual cleaning of the contacting surfaces takes
place during the experiment: the tip is crashed forcibly and repeatedly into
the substrate causing extended plastic deformation until a good metallic
contact is obtained. Pressing tip and substrate hard has the effect not only
of displacing adsorbates or oxide on the surface but also of involving large
amounts of material in the formation process which makes possible the
fabrication of longer necks. This fabrication procedure contrasts with
the {\em gentle} contact formation  by other authors both in
conductance \cite{Pascual,Olesen} and force \cite{Durig} measurements.

In the STM experiments formation of a good metallic contact is easily
recognized because after rupture the apparent tunneling barrier attains a
high value (3-4 eV) and separation of tip and substrate results in the
formation of a protrusion on the substrate that can be imaged. In this case
the last contact has been shown to be of the order of $2e^{2}/h$. Formation
of a good metallic contact is even easier to recognize in the STM/AFM
experiments since contamination produces a repulsive force while in the
tunneling regime.

After the initial preparation of the contact described above,
repeated cycles of elongation and contraction with a given amplitude
typically result in a very similar current
(and force, in the STM/AFM experiments)
evolution, that is, a kind of steady state is reached.
We must remark that the system remains in this steady state until
the amplitude or offset of the cycles is altered, and in some cases
as we will see below subsequent cycles are very repetitive.
In a marked contrast
with macroscopic plastic deformation the contact shows no sign of fatigue.
We only present results for
these steady-state cycles.
The usual data acquisition parameters are 2048
data points per cycle at a rate of 10-100 $\mu $s, and typically sets of
five cycles are acquired.

It is important to note that the necks (slope and hysteresis of the cycles)
depend strongly on the fabrication procedure. Gentle contact causes necks of type A
(see Fig.\ \ref{fig.curves}) while repeated cycling produces necks of type D.

\subsection{Shape of the necks}

\begin{figure}
\vspace{0mm}
\begin{center}
\leavevmode
    \epsfxsize=70mm
 \epsfbox{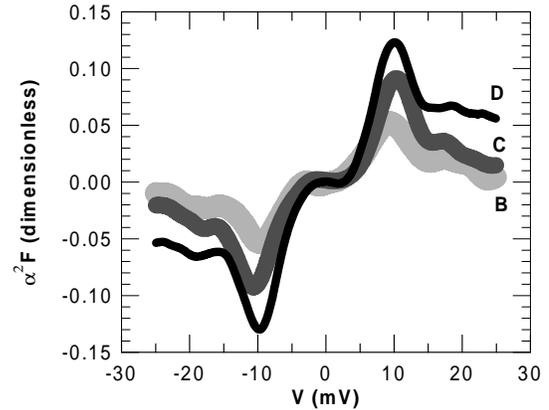}
 \end{center}
\vspace{0in}
\caption[]{PCS curves at 4.2 K for  Au necks whose behaviors are similar to those in
        Fig.\ \ref{fig.curves}.
        The black curve corresponds to a cycle
        of type D, at a conductance of 83 units;
        dark gray curve corresponds to type C, at conductances 77, 106
        and 130 conductance units; light gray curve corresponds to type B
        at conductances 77, 103, 130 conductance units.
        All the curves for a given cycle are included within thickness of the
        plotted curves.}
\label{fig.phonon}
\end{figure}

In order to determine the minimal cross section from the conductance using
the modified Sharvin formula eq.\ \ref{eq.Sharvin}, it is necessary to be
certain that the necks are ballistic. For the experiments performed at low
temperature we have used point-contact spectroscopy (PCS). Fig.\ \ref
{fig.phonon} shows the PCS curves corresponding to necks, that is steady
state plastic deformation cycles, similar to D, C,
and B in Fig.\ref{fig.curves}. For a given neck, the
PCS curve does not change with the conductance, that is the derivative of
the conductance scales with $R^{3/2}$ as given by eq.\ \ref{eq.spectral}, but
the amplitudes of the phonon peaks are different for different necks. For a
neck such as D, the amplitude of the phonon peaks is maximum (black curve in
Fig.\ \ref{fig.phonon}) and similar to that of previously reported spectra
\cite{Jansen2} for ballistic point contacts, indicating that these necks are
quite crystalline \cite{Ralph}. In necks such as B and C the amplitude of
the phonon peaks in the PCS curve is reduced (gray curves in Fig.\ \ref
{fig.phonon}) due to elastic scattering in the neck region, indicating the
presence of defects, but still the necks are far from being disordered. From
this evidence we conclude that our experimental metallic necks formed by
plastic deformation are quite ballistic and not disordered. This idea is
supported by the results of the molecular dynamics simulations \cite
{Landman,Lynden-Bell}.

In Fig.\ \ref{fig.curves} we show four representative experimental sets of
data. Each set consists of 5 elongation-contraction cycles. In order to use
the previously defined slab model to deduce the shape of the corresponding
necks, we have to find the equilibrium points (zero elastic force points) of
each configuration, which is approximately in the middle of the hysteresis
cycle as illustrated in Fig.\ \ref{fig.forcemodel}. The {\em plastic
deformation length} $\lambda$ for these necks
at the largest radius (approximately 1.9 nm,
corresponding to 120 conductance units) is: for curve A, 0.2 nm; for curve
B, 1 nm; for curve C, 2.8 nm; and for curve D, about 6 nm. In all cases,
except for curve A, this value decreases monotonically. Taking into account
the physical meaning of $\lambda$, we can see that for neck D
the amount of material involved in the plastic deformation
is many times larger than for neck A. In
this last case only one atomic layer is involved while in the former case
many layers redistribute in the process of plastic deformation.

\begin{figure}
\vspace{0mm}
\begin{center}
\leavevmode
    \epsfxsize=70mm
 \epsfbox{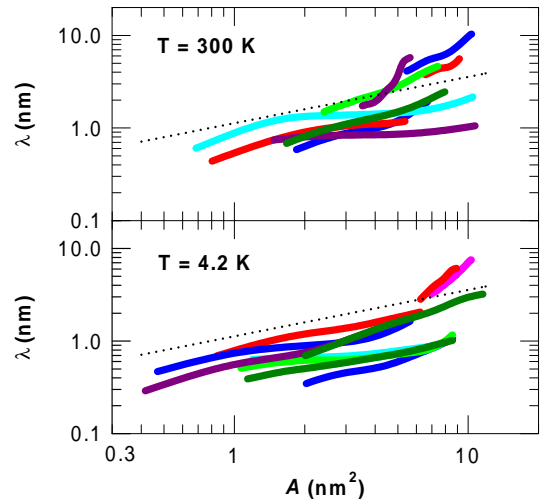}
 \end{center}
\vspace{0in}
\caption[]{Plastic deformation length {\em vs} minimal cross section
of  Au necks fabricated at room temperature and 4.2 K. The dashed line represents
$\lambda=2\sqrt{A/\pi}$, {\em i.e.,} an aspect ratio of one.}
\label{fig.loglog}
\end{figure}


In order to summarize all the different observed behaviors, in Fig.\ \ref
{fig.loglog} we show a log-log plot of $\lambda$ {\em vs} $A$
at room temperature and at 4.2 K. The few
curves plotted are representative of the observed behavior for hundreds of
necks of cycles. All the observed curves are within the same region in the
plot. The dashed line corresponds to a $\lambda$ equal to the diameter of the
neck. Note that some curves lie above this line, that is, the longest necks
are longer than their diameter.
Naturally $\lambda$ decreases to a few
tenths of nanometer as the minimal section decreases to that corresponding
to a few atoms.
On the other hand, the shortest necks show
small variations of $\lambda$, typically from 0.5 to 1 nm.
Necks formed at low and room temperature are not very different
(at room temperature long necks are somewhat easier to form
and typically  about 30\% longer). This is to be expected since room temperature
is still much  lower than the melting temperature for Au.

\begin{figure}[h]
\vspace{0mm}
\begin{center}
\leavevmode
    \epsfxsize=70mm
 \epsfbox{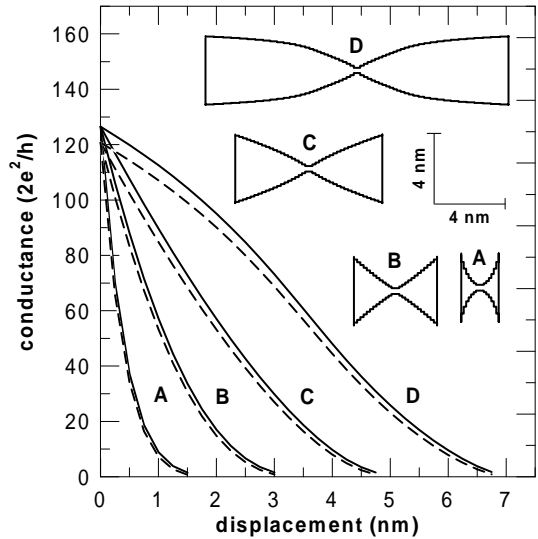}
 \end{center}
\vspace{0in}
\caption[]{Conductance and shapes of the necks in Fig.\ \ref{fig.curves}
using slab model.}
\label{fig.necks}
\end{figure}

\begin{figure}
\vspace{0mm}
\begin{center}
\leavevmode
    \epsfxsize=70mm
 \epsfbox{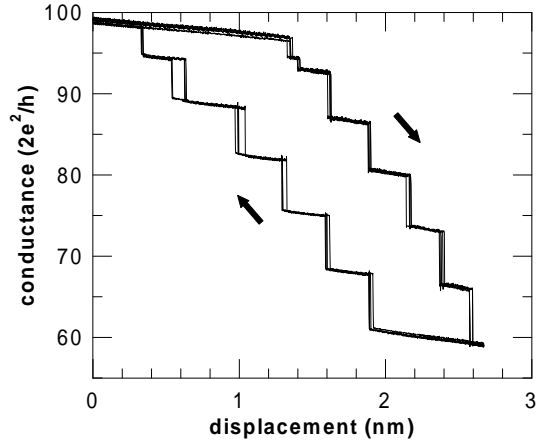}
 \end{center}
\vspace{0in}

\caption[]{Striking repetitiveness of the conductance during the cyclical
plastic deformation of a long Au neck at 4.2 K.
This is a set of five consecutive elongation-contraction cycles. The neck
goes through the same (or very similar) configurations during elongation and
contraction in all the cycles.}
\label{fig.goodneck}
\end{figure}

In Fig.\ \ref{fig.necks} we show the geometry of four necks right before
breaking. We have chosen the dependence of $\lambda$ on $A$ similar to that
corresponding to the experimental necks in Fig.\ \ref{fig.curves}. This
figure illustrates the fact that starting from the same cross sectional area
it is possible to have necks with very different geometries. We must
emphasize that necks labelled A and D are the limiting cases. Neck A
{\em (short neck)} has a
constant $\lambda\approx 0.2-0.3$ nm. On
the other hand, neck D {\em (long neck or nanowire)}
corresponds to the longest observed necks.
Note that according to the model the shape of the neck
is only determined by $\lambda(A)$ and not by the initial cross sectional
area of the neck. Neck D not only is longer but also, as shown by the PCS
curve, it has less defects, that is, it is more crystalline. Another
interesting features of these long necks is that their
evolution is typically very repetitive.
Fig.\ \ref{fig.goodneck} shows another long neck. The
repetitiveness is remarkable, not only the five consecutive cycles superpose
almost perfectly, but the configurations appear to be the same (or very
similar) for elongation and contraction. These long necks very often evolve
repetitively for an indefinite number of cycles once the steady state is
reached. This repetitiveness is not observed in short necks. This behavior
is likely to be related to the deformation mechanism during plastic
deformation. For short necks, relaxation would involve structural
transformation in which a portion of the neck of thickness $\lambda$
composed of just a few atomic layers disorders and then rearranges to form
an added layer \cite{Landman,MoleSlab}, that is, some kind of order-disorder
transition. This disorder precludes
repetitiveness. On the other hand, the values of $\lambda$ for long necks
indicate that the structural transformation involves many atomic layers. In
this case the deformation mechanism is likely to involve gliding on (111)
planes (as illustrated by the recent molecular dynamics simulations in Ref.\
\onlinecite{Landman2,pep}) instead of disorder and rearrangement \cite{Landman},
making possible a cyclic evolution through almost identical configurations.

\begin{figure}[h]
\vspace{0mm}
\begin{center}
\leavevmode
    \epsfxsize=70mm
 \epsfbox{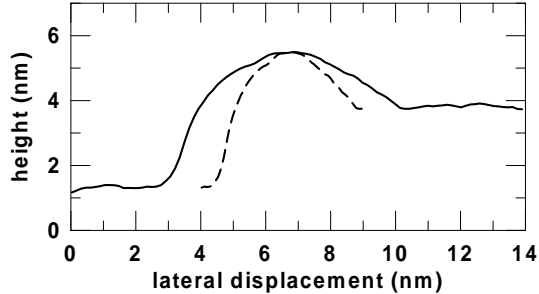}
 \end{center}
\vspace{0in}

\caption[]{Profile of a protrusion resulting after breaking a
long Au neck ($T=4.2$ K).
If we assume that the protrusion on the substrate is scanned with a
similar protrusion acting as STM tip the actual dimensions of the protrusion
would be given approximately by the dashed line.}
\label{fig.image}
\end{figure}

If the substrate is imaged right after breaking a neck
a protrusion is observed on the spot where
the neck was formed. Fig.\ \ref{fig.image} shows the profile of
such a protrusion.
The observed profile must be corrected (dashed line)
taking into account that a similar protrusion is acting as
scanning tip. These dimensions are consistent with the shapes
estimated using the slab model.

The geometries of the necks in Fig.\ \ref{fig.necks} suggest that when a
neck of type A breaks at low  temperatures when surface difussion is negligible
the last few atoms should form a sharp spire.
However,
this kind of structure is unlikely to be stable due to surface tension.
This final sharp structure retracts several tenths of nanometer and
only a blunt protrusion is visible.

The different types of necks described above depend on the history of
the contact. In the case of Au illustrated here, short necks (type A)
are typically
obtained for the initial indentations, while long necks (type D)
require deeper
indentation and repeated elongation and contraction. The fact that long
necks present a prominent and repetitive stepped structure in the
conductance (like the neck in Fig.\ \ref{fig.goodneck}), and a large
amplitude of the phonon peaks in the PCS curve, indicates that the neck is
quite crystalline. Possibly this repeated plastic deformation causes a
``mechanical annealing'' of the defects.

\section{Conclusions}

A simple slab model that
correlates minimal cross-sectional
areas determined from the electrical conductance during
the contraction and elongation process
to the relative tip-substrate displacement
can be used to estimate the
shape of the metallic neck formed using an STM,
and point contact spectroscopy (PCS)
can be  used to check that the constriction is indeed
ballistic which is
essential for the correct determination of the minimal cross section.

Determination of the shape of experimental Au necks at
low (4.2 K) and room temperatures shows that there
is a strong dependence on the fabrication procedure, being it possible to
form long necks (or nanowires) by repeated elongation and contraction,
while necks formed at initial shallow indentations are typically short.

The striking repetitiveness of the conductance
during elongation-contraction
cycles of the long necks and the fact that
electronic transport is ballistic
as indicated by PCS measurements shows that long necks
quite crystalline.

\acknowledgments

This work was supported by the DGICYT under contracts
MAT95-1542 and  PB94-0382.


\begin{references}
\bibitem{Muller}  C.J. Muller, J.M. van Ruitenbeek and L.J. de Jongh, Phys.
Rev. Lett. {\bf 69}, 140 (1992); J.M. Krans, J.C. Muller, I.K. Yanson,
Th.C.M. Govaert, R. Hesper, and J.M. van Ruitenbeek, Phys. Rev. B {\bf 48},
14721 (1993); N. van der Post, E.T. Peters, I.K. Yanson, and J.M. van
Ruitenbeek, Phys. Rev. Lett. {\bf 73}, 2611 (1994); J.M. Krans and J.M. van
Ruitenbeek, Phys. Rev. B {\bf 50}, 17659 (1994);
J.M. Krans, J.M. van Ruitenbeek, V.V. Fisun, I.K. Yanson, and L.J. de Jongh,
Nature (London) {\bf 375}, 767 (1995).

\bibitem{Agrait}  N. Agra\"\i t, J.G. Rodrigo and S. Vieira, Phys. Rev. B
{\bf 47}, 12345 (1993); N. Agra\"\i t, J.G. Rodrigo, C. Sirvent and S.
Vieira, Phys. Rev. B {\bf 48}, 8499 (1993).

\bibitem{Rodrigo}  J.G. Rodrigo, N. Agra\"\i t, and S. Vieira, Phys.\ Rev.\
B {\bf 50}, 374 (1994).

\bibitem{Rodrigo2}  J.G. Rodrigo, N. Agra\"\i t, C. Sirvent, and S. Vieira,
Phys.\ Rev.\ B {\bf 50}, 7177 (1994); J.G. Rodrigo, N. Agra\"\i t, C.
Sirvent, and S. Vieira, Phys.\ Rev.\ B {\bf 50}, 12788 (1994).

\bibitem{Smith}  D.P.E. Smith, Science {\bf 269}, 371 (1995).

\bibitem{Pascual}  J.I. Pascual, J. M\'endez, J. G\'omez-Herrero, A.M.
Bar\'o, N. Garc\'\i a, and Vu Thien Binh, Phys.\ Rev.\ Lett.\ {\bf 71}, 1852
(1993). J.I. Pascual, J. M\'endez, J. G\'omez-Herrero, A.M. Bar\'o, N.
Garc\'\i a, U. Landman, W.D. Luedtke, E.N. Bogachek, and H.-P. Cheng,
Science {\bf 267}, 1793 (1995).

\bibitem{Olesen}  L. Olesen, E. L\ae gsgaard, I. Stensgaard, F. Besenbacher,
J. Schi\o tz, P. Stolze, K.W. Jacobsen, and J.K. N\o rskov, Phys.\ Rev.\
Lett. {\bf 72}, 2251 (1994).


\bibitem{PRL1}  N. Agra\"\i t, J. G. Rodrigo, G. Rubio, C. Sirvent and
S.Vieira, Thin Solid Films {\bf 253}, 199, (1994);
N. Agra\"\i t, G. Rubio, S. Vieira,
Langmuir {\bf 12}, 4505 (1996);
N. Agra\"\i t, G. Rubio,
S. Vieira, Phys. Rev. Lett. {\bf 74}, 3995 (1995).

\bibitem{Durig}  A. Stalder and U. D\"urig, Appl. Phys. Lett. {\bf 68}, 637
(1996); J. Vac. Sci. Technol. B {\bf 14}, 1259 (1996)

\bibitem{PRL2}  G. Rubio, N. Agra\"\i t, S. Vieira, Phys. Rev. Lett. {\bf 76}%
, 2302 (1996).

\bibitem{Landman}  U. Landman, W.D. Luedtke, N.A. Burnham, and R.J. Colton,
Science {\bf 248}, 454 (1990).

\bibitem{Lynden-Bell}  R. M. Lynden-Bell, J. Phys.\ Condens.\ Matter {\bf 4}%
, 2127 (1992); Science {\bf 263}, 1704 (1994).

\bibitem{Landman2}  U. Landman, W.D. Luedtke, B.E. Salisbury, R.L. Whetten,
Phys.\ Rev.\ Lett.\ {\bf 77}, 1362 (1996).

\bibitem{Todorov}  T.N. Todorov and A.P. Sutton, Phys.\ Rev.\ Lett.\ {\bf 70}%
, 2138 (1993); A.M. Bratkovsky, A.P. Sutton and T.N. Todorov, Phys.\ Rev.\ B
{\bf 52}, 5036 (1995).

\bibitem{Bogachek}  E.N. Bogachek, A.M. Zagoskin and I.O. Kulik, Sov. J. Low
Temp. Phys. {\bf 16}, 796 (1990).

\bibitem{Torres1}  J.A.Torres, J.I. Pascual and J.J. S\'aenz, Phys.\ Rev.\ B
{\bf 49}, 16581 (1994).

\bibitem{Torres2}  J.A. Torres and J.J. S\'aenz, Physica B {\bf 218}, 234
(1996).

\bibitem{Bratkovsky}  A. M. Bratkovsky, and S. N. Rashkeev, Phys. Rev. B
{\bf 53}, April (1996).

\bibitem{Gimzewski}  J.K. Gimzewski, R. M\"oller, Phys.\ Rev.\ B {\bf 36},
1284 (1987).

\bibitem{mfpAu}  The mean free path of the electrons in Au is about 30 nm at
room temperature and much larger at low temperature.

\bibitem{Sharvin}  Yu.V. Sharvin, Zh. Eksp. Teor. Fiz. {\bf 48}, 984 (1965)
[Sov. Phys. JETP {\bf 21}, 655 (1965)].

\bibitem{Jansen}  A.G.M. Jansen, A.P. van Gelder, and P. Wyder, J. Phys. C:
Solid St. Phys., {\bf 13}, 6073 (1980).

\bibitem{Mole}  A. Garc\'\i a-Mart\'\i n, J.A. Torres and J.J. S\'aenz,
Phys.\ Rev. B {\bf 54}, (in press), 1996.

\bibitem{Serena}  P. Garc\'\i a-Mochales, P.A. Serena, N. Garc\'\i a, J.L.
Costa-Kr\"amer, Phys.\ Rev.\ B {\bf 53}, 10268 (1996).

\bibitem{Ralph}  I.K. Yanson, Fiz. Nizk. Temp. {\bf 9}, 676 (1983) [Sov. J.\
Low Temp.\ Phys. {\bf 9}, 343 (1983)];
A.M. Duif, A.G.M. Jansen and P. Wyder, J.\ Phys.: Condens.\
Matter {\bf 1}, 3157 (1989);
see also D.C. Ralph and R.A. Buhrman,
Phys.\ Rev.\ B {\bf 51}, 3554 (1995).

\bibitem{MoleSlab}  J.A. Torres, J.J. Sa\'enz, Phys.\ Rev.\ Lett.\ {\bf 77},
2245 (1996).

\bibitem{Jansen2}  A.G.M. Jansen, F.M. Mueller and P. Wyder, Phys.\ Rev.\ B
{\bf 16}, 1325 (1977).

\bibitem{pep} P. Espa\~{n}ol, I. Zu\~{n}iga, G. Rubio, N. Agra\"{\i}t,
to be published.


\end{references}
\end{document}